\documentclass[12pt]{article}
\usepackage{graphicx}
\def\bra#1{\mathinner{\langle{#1}|}}
\def\ket#1{\mathinner{|{#1}\rangle}}
\begin{document}

\begin{flushright}
McGill 06/05 \\
\end{flushright}

\begin{center}
\bigskip
{\Large \bf \boldmath
Compton effect: interacting particles or interacting waves.
} \\
\bigskip
\bigskip
{\large
Oscar F. Hern\'andez$^{\small \cite{marianopolis}}$ }
\\
\vskip2mm
oscarh@physics.mcgill.ca \\
\vskip1mm
 {\it Physics Department, McGill University,
\\ 3600 University St.,
\\ Montr\'eal, Qu\'ebec, Canada,
H3A 2T8.}\\
\end{center}

\begin{center}
\bigskip
%
\parbox[t] {\textwidth}
{\it {\bf Abstract:} Traditional textbook explanations of the
Compton effect treat the photon electron interaction as a particle
collision. This explanation is a pedagogical disaster, implying that
sometimes interactions are particle-like whereas quantum mechanics
always demands that they be wave-like; a photon wavefunction evolves
according to a wave equation until its collapse at measurement. If
this is so why then does the classical radiation wave equation fail
to predict the Compton effect? We address these issues and propose a
clearer explanation.}
\end{center}

\baselineskip=14pt


Traditional quantum mechanics textbook present the Compton effect as
proof that the photon is a particle that carries momentum. The
correct value of the Compton shift to longer wavelength is derived
by applying relativistic two-body kinematics with the photon and
electron treated as particles. This traditional explanation states
wavelenght shift occurs from "individual photons scattering
elastically off individual electrons"$^{\small \cite{gasiorowicz}}$.
That the photon and the electron must be
treated as particles when they scatter from each other means that
any interaction between a photon and an electron necessarily
collapse their wavefunctions. Yet X-ray diffraction treats photons
as waves that interfere even after scattering from atoms' electrons.
Their wavefunctions only collapse after they have been measured at a
photographic plate.

In fact since no measurement occurs during the collision, the
Compton effect's photon electron interaction should be analysed as
the scattering of photon plane waves from an (approximately
stationary) electron. Since both classical electromagnetism and
quantum mechanics are wave theories that conserve energy and
momentum one is led to wonder why classical radiation theory does
not correctly predict the Compton effect. Classical electromagnetism
is Lorentz invariant and electromagnetic waves do carry momentum and
energy through the Poynting vector. With this reasoning it is
puzzling why one does not derive a Compton-like effect if one treats
the electron as a relativistic point particle scattering radiation.
The reason is that in classical radiation the missing
energy-momentum taken up by the recoiling electron is distributed
throughout the entire wave. When the wave reaches our detector it
all carries part of the entire energy-momentum. However in a quantum
theory, this is where a measurement is taking place. In other words
this is where a low intensity wave collapses into one particle. The
entire missing energy and momentum must now be carried by this one
photon since the rest of the wavefunction is now zero in the rest of
the $4\pi$ solid angle encompassed by our detector.

This explanation is very different from the traditional one where
the photon's wavefunction collapses at the interaction point not at
the detector. The traditional explanation gives the impression that
sometimes interactions take place as particles and sometimes as
waves. Which point of view to take remains a mystery. The
explanation proposed here removes the mystery from it all. In the
quantum world all interactions happen as waves. Particle
interpretations occur only in the detector.

If this proposed explanation is correct we should be able to
pinpoint the quantum explanation where the observed particle number
enters in. If this number is one we recover the Compton effect. If
this number is taken to infinity we recover classical radiation
theory, i.e. Thompson scattering.

We begin by considering the scattering of two wave packets.
Regardless of whether we are working within quantum mechanics or a
classical wave theory such as electromagnetism, the definition of
the differential cross section is the same. Detectors in both cases
detect energy. The energy deposited in the detector can be
interpreted as a wave intensity or as a particle's quanta of energy.
Energy flow can refer to classical energy flow or particle number
flow. The flow through a surface is given by the relevant current
conservation equation; the continuity equation in classical
electromagnetism or the probability conservation equation in quantum
mechanics. The scattered energy into a particular solid angle per
unit time, is proportional to the incident energy passing through a
unit area per unit time. The proportionality constant is defined as
the differential cross section:

\begin{equation}\label{crosssection}
\frac{{d\sigma }}{{d\phi d\Omega }} =
\frac{{\rm [scattered\ energy]  }
/ {\rm [(solid\ angle)(time)]}}
{ {\rm [incident\ energy]    } /  {\rm[(area)(time)] }  }
\end{equation}

This cross section formula can be factorized into two parts; a
density into final states part $\rho(E)$, and a transition matrix
element part $|\bra{\alpha}T\ket{\beta}|^2$. In quantum mechanics
this factorized formula goes by the name of Fermi's Golden Rule, but
it's validity is beyond quantum mechanics and is true for all wave
phenomena, be they quantum waves or classical waves.

\begin{equation}\label{factorize}
\frac{{d\sigma }}{{d\phi d\Omega }} =
\rho(E)\times |\bra{\alpha}T\ket{\beta}|^2
\end{equation}

The density into final states includes the number of particles we
are trying to detect and their kinematics. The transition matrix
element is what is calculated in time dependent perturbation theory.
This contains the dynamical details of the interactions. The
important point is that the Compton effect depends on the density of
states, i.e. the kinematics, not the dynamical details of the
interaction. It is the density of final states that will change
depending on whether we are trying to detect one photon versus many
(i.e. a wave). If the density of states is for one photon and one
electron we will recover the Klein-Nishina formula for the Compton
effect. If the density of states is for many many photons and one
electron, we recover Thompson formula for the scattering of
radiation by a free electron. In other words the quantum mechanical
wave representing many photons can now be interpreted as a classical
wave.

It is easy to understand why the many particle limit is equivalent
to the low energy limit usually quoted in texts as a way of
recovering Thompson scattering from Compton scattering. The
kinematics in the density of states depends on the number of final
particles. The initial scattering is always represented by two wave
packets. Hence the kinematics of energy-momentum conservation will
always be equivalent to that of two particles (a photon and an
electron) scattering into $N$ final particles. If $N$ is large then
the particles individual energy must necessarily be small since the
initial energy remains constant.


\bigskip
\noindent {\bf Acknowledgements}:
I would like to thank Guy Moore for valuable discussions and Jim Cline for
proofreading the text.


\end{document}